\documentclass[preprint,showpacs,preprintnumbers,amsmath,amssymb]{revtex4}

\usepackage{bm}
\usepackage{epsfig}

\begin{document}


\title{Singe ferroelectric and chiral magnetic domain of 
single-crystalline BiFeO$_3$ in an electric field}

\author{Seongsu Lee,$^1$ Taekjib Choi,$^1$ W. Ratcliff II,$^2$ R. 
Erwin,$^2$ S-W. Cheong,$^1$ and V. Kiryukhin$^1$}

\affiliation{$^1$Rutgers Center for Emergent Materials and Department of 
Physics and Astronomy, Rutgers University, Piscataway, NJ 08854, USA}

\affiliation{$^2$NIST Center for Neutron Research, NIST, Gaithersburg, MD 
20899, USA}

\date{\today}%

\begin{abstract} 

We report polarized neutron scattering and piezoresponse force microscopy 
studies of millimeter-sized single crystals of multiferroic BiFeO$_3$. The 
crystals, grown below the Curie temperature, consist of a single 
ferroelectric domain. Two unique electric polarization directions, as well 
as the populations of equivalent spiral magnetic domains, can be switched 
reversibly by an electric field. A ferroelectric monodomain with a 
single-$q$ single-helicity spin spiral can be obtained. This level of 
control, so far unachievable in thin films, makes single-crystal BiFeO$_3$ 
a promising object for multiferroics research.

\end{abstract}

\pacs{77.80.-e,77.80.Dj,77.80.Fm,75.80.+q} 
\maketitle

Understanding and control of domain structure is crucial for applications 
of magnetic and ferroelectric materials, as well as for studies of their 
intrinsic properties. Coercivity of permanent magnets, switching of memory 
devices, and many other functional properties are defined by domain 
physics. Ferroelectricity and magnetism coexist in 
multiferroics\cite{Cheong}. These materials have attracted significant 
attention recently because of an intriguing possibility of control of 
their magnetic properties with an electric field and vice 
versa,\cite{Zhao,Eerenstein,Ramesh} leading to discovery of new physical 
phenomena and fabrication of novel devices\cite{Chu}. So far, little 
research has been done on the domain structure of multiferroics. Among a 
few notable exceptions\cite{Aken} is BiFeO$_3$ (BFO), but mostly in its 
thin film form, since suitable single crystals were unavailable. BFO is 
arguably the most extensively studied multiferroic due to its large 
polarization at room temperature\cite{Ramesh}. Importantly, this is the 
only multiferroic exhibiting large effects of an electric field on the 
magnetic structure, as first demonstrated in thin films\cite{Zhao}, and 
subsequently in single crystals\cite{Lee,Lebeugle}. Unfortunately, BFO 
thin films exhibit poor crystallinity and are influenced by substrate 
strain,\cite{Zhao,Chu} impeding studies of BFO's intrinsic properties, and 
complicating device fabrication. Herein, we report studies of 
single-crystal BFO in which an unprecedented level of control of domains 
and their electric and magnetic properties can be achieved.

BFO is ferroelectric for $T$$<$$T_C$$\approx$850 ${}^\circ$C. It exhibits 
the rhombohedral $R3c$ structure with a perovskite pseudocubic unit cell 
($a$$\approx$3.96 \AA, $\alpha$$\approx$89.4${}^\circ$) elongated along 
the (111) direction which coincides with the electric polarization vector 
$\bf{P}$.\cite{Sos1} For $T$$<$$T_N$$\approx$370 ${}^\circ$C, Fe spins 
exhibit the $G$-type antiferromagnetic order \cite{Moreau}. In single 
crystals, there is an additional long-range ($\lambda$$\approx$620 \AA) 
magnetic modulation with the wavevector $\tau_1$=($\delta$,0,-$\delta$), 
believed to be cycloidal with antiparallel spins slowly rotating in the 
plane containing $\bf{P}$ and $\bf{\tau}$.\cite{Sos1} Wavevectors 
$\tau_2$=($\delta$,-$\delta$,0) and $\tau_3$=(0,-$\delta$,$\delta$) are 
equivalent by symmetry, giving three propagation directions, each 
perpendicular to $\bf{P}$. The spins are rigidly coupled to the crystal 
lattice, and rotation of $\bf{P}$ between the different cubic body 
diagonals by an electric field results in the rotation of the magnetic 
structure.\cite{Zhao,Lee,Lebeugle} This effect was recently used to 
demonstrate local control of magnetization by an electric field.\cite{Chu}

BFO exhibits four structural variants (ferroelastic domains) based on the 
four different cubic body diagonals, and eight possible directions of 
$\bf{P}$ (ferroelectric domains) shown in Fig. 1(a). To produce any 
realistic device, and to study the intrinsic properties, single-domain 
samples are needed. However, despite significant efforts, the best 
available BFO thin films always consist of $\mu$m-sized ferroelastic 
domains.\cite{Zhao,Chu} Films with suitable electrical characteristics are 
of the (001) type, and therefore poling (electric field applied along the 
cubic direction) is inappropriate for getting rid of the ferroelastic 
domains. In single crystals, poling in the (111) direction might be 
possible, but the situation is complicated by the presence of the 
equivalent magnetic domains in a single ferroelectric domain, as shown in 
Fig. 1(b). In this work, we report preparation of mm-sized single crystals 
with a single ferroelectric and single-$\tau$ magnetic domain, see Fig. 
1(c). The magnetic state is chiral, as in a single-helicity spiral. The 
electric and magnetic state can be reversibly controlled by an electric 
field, which switches the electric polarization between the two unique 
states shown in Fig 1(d), and also controls the equivalent magnetic 
domains. These results open new opportunities for studies of BFO, and for 
BFO-based device fabrication.

Single crystals of BFO were grown using a 
Bi$_2$O$_3$/Fe$_2$O$_3$/B$_2$O$_3$ flux by slow cooling from 870 to 620 
${}^\circ$C. Crystals with $\sim$mm$^2$-area natural faces normal to the 
cubic directions were obtained. Piezoresponse force microscopy (PFM) was 
performed with a commercial scanning probe microscope. $P_x$, $P_y$, and 
$P_z$ components of $\bf{P}$ were measured, with $P_z$ normal to the 
sample surface, and the axes coinciding with the pseudocubic directions. 
For electric domain switching, a 40 V bias (bottom electrode) was applied 
to a 20 $\mu$m-thick sample, with the PFM tip grounded. Neutron 
experiments were done at $T$=50 K on BT9 and BT7 (polarized measurements) 
triple-axis spectrometers at NIST Center for Neutron Research at an energy 
of 14.7 meV with collimations 40-10-s-40-80 (BT9) and 25-10-s-50 (BT7). 
The collimations were set such that their further tightening would result 
either in no significant resolution improvement, or in unacceptable signal 
losses, ensuring the best possible resolution compatible with our signal 
level. Two parallel gold contacts were evaporated on the largest faces of 
a 3$\times$2$\times$1 mm$^3$ sample for an electric field $E$ to be 
applied in the (001) direction.

Different ferroelastic domains can be easily detected using our 
neutron scattering setup \cite{Lee}. The only Bragg peaks found in our 
samples were due to a single domain, even though a careful search for 
other domains was done. Considering the experimental accuracy, we 
conclude that more than 95\% of each studied sample consist of a single 
ferroelastic domain. Multiple PFM measurements show the same $P_x$, 
$P_y$, and $P_z$ across the surface in the majority of 
as-prepared samples. Typical images are shown in Fig. 2(a-c). Thus, the 
samples consist of a single ferroelectric domain. Local electric 
polarization can be switched reversibly by application of $E$ 
normal to the (001) surface by the PFM tip. Fig. 2(d-f) 
show that while $P_z$ changes its direction in the region of bias 
application, $P_x$ and $P_y$ remain unchanged in the process. 
Thus, $\bf{P}$ rotates by 71$^\circ$ as shown in Figs. 1(d) 
and 2(g), the rotation direction is unique, and no 109$^\circ$ or 
180$^\circ$ domains are created. This observation is in agreement with 
theoretical calculations showing that the elastic energy losses are 
minimal for the 71$^\circ$ domain boundaries.\cite{Streiffer}

We now turn to the magnetic structure of the equivalent magnetic domains 
$\tau_1$, $\tau_2$, and $\tau_3$ in the ferroelectric domain $\bf{P}_A$ of 
Fig. 1(b). Fig. 3 shows unpolarized neutron scattering pattern near the 
(1/2,-1/2,1/2) $\pm$$\bf{\tau}$ magnetic peak in the ($h$,-$h$,$l$) plane, 
with projections of all the three magnetic wavevectors marked. Only the 
$\bf{P}_A$-domain signal is detected in this geometry, even when the 
second ferroelectric domain ($\bf{P}_B$ in Fig. 1) starts appearing for 
$E$$<$0. (See Ref. \cite{Lee} for the discussion of the $\bf{P}_B$-domain 
magnetism.) For $E$=0, two magnetic domains, with wavevectors $\tau_1$ and 
$\tau_2$, are present. Application of $E$ at an acute angle of 55$^\circ$ 
to $\bf{P}$, as shown in Fig. 1(c), suppresses domain $\tau_2$. For 
$E$=1.3 MV/m, only domain $\tau_1$ is observed within the accuracy of our 
measurement ($\sim$10\% of the domain population), see Fig. 3(b). This 
figure shows that a fit to two Gaussian peaks at $\pm\tau_1$, with the 
widths and orientation closely matching the calculated resolution, 
reproduces the data very well. Note that for the out-of-scattering-plane 
vector $\tau_1$, the peak positions differ slightly from their projections 
on the scattering plane due to resolution effects. Thus, the entire 
crystal consists of a single-$\bf{\tau}$ magnetic domain, as shown 
schematically in Fig. 1(c). Sufficiently large negative $E$ produces the 
opposite effect, enriching domain $\tau_2$ at the expense of $\tau_1$, see 
Fig. 3(c). This effect is probably driven by the uniaxial piezoelectric 
strain which is linearly coupled to $E$. Figs. 1(e,f) show that the field 
projection on the plain containing $\tau_1$, $\tau_2$, and $\tau_3$ is 
normal to $\tau_2$, and therefore the induced strain lifts the symmetry 
making the domains equivalent in zero field. Note that $\tau_2$ is the 
only common magnetic wavevector for $\bf{P}_A$ and $\bf{P}_B$ domains, 
which may play a role in its stabilization for $E$$<$0. The absence of 
domain $\tau_3$ is probably related to the post-growth residual strain. To 
characterize the magnetic domain populations, the neutron data were 
modeled by four Gaussian peaks of the same shape (measured at $E$=1.3 
MV/m), see Fig. 3(a-c). For $\tau_1$, the peaks were located in the 
experimental positions found at $E$=1.3 MV/m. For the in-plane $\tau_2$, 
the peaks were located in the theoretical positions. As functions of $E$, 
the domain populations form reproducible hysteresis loops, showing 
electric field control of the magnetic domains, see Fig. 3(d). While 
higher-resolution measurements are needed to reveal the details of these 
loops, our data strongly suggest that any specific magnetic domain can be 
stabilized by application of the appropriate uniaxial stress, and that 
complete control of the magnetic state can thus be achieved in BFO.

Cycloidal magnetic state is believed to be realized in BFO,\cite{Sos1} but 
alternatives were also proposed.\cite{Przenioslo} While recent 
single-crystal neutron studies support the cycloid,\cite{Lebeugle} 
non-collinear states are difficult to analyze, and further work is needed 
to establish the magnetic state unambiguously. Various magnetic 
interactions can produce the cycloid. Common cases include competing 
nearest and next nearest neighbor exchange, and antisymmetric 
Dzyaloshinsky-Moria (DM) and magnetoelectric \cite{Kadomtseva} 
interactions.\cite{Cheong} In the (1-10) FeO chains in ferroelectric BFO, 
all the exchange-mediating oxygens are shifted from symmetric positions 
with respect to Fe in the same direction along $\bf{P}$.\cite{Sos1} For 
such shifts, the magnetoelectric interaction is proportional to 
P$(m_yL_x-m_xL_y)$, where $m$ is the unit magnetization, $L$ is the 
antiferromagnetic vector, and the $xy$-plane is defined as normal to the 
direction of $\bf{P}$.\cite{Kadomtseva} It is of the same form as the DM 
interaction, and should produce a weakly ferromagnetic 
spiral\cite{Kadomtseva,Sos2} with rotation direction (helicity) uniquely 
defined in the rotation plane by $\bf{P}\sim \bf{e}\times\bf{\tau}$, where 
$\bf{e}$ is the spin rotation axis,\cite{Cheong} see insets in Fig. 4. 
Non-zero $\bf{P}$ is essential for the chiral state, \textit{i.e}. the 
spiral is induced by $\bf{P}$ which appears at $T$ much higher than $T_N$. 
(Note that this is different from magnetically-driven ferroelectrics such 
as RMnO$_3$, in which the situation is reversed, and $\bf{P}$ is induced 
by the inverse DM interaction in a magnetic spiral.\cite{Cheong}) We used 
polarized neutron scattering to characterize the helicity of the spiral. 
Scans were taken in both of the spin-flip channels, denoted as (+-) and 
(-+), with neutrons polarized parallel to the scattering vector $\bf{Q}$. 
Scattering cross section\cite{Lovesey} is proportional to $\langle 
S_{\perp Y}S_{\perp Y}^*\rangle+\langle S_{\perp Z}S_{\perp Z}^*\rangle\pm 
i(\langle S_{\perp Y}S_{\perp Z}^*\rangle- \langle S_{\perp Z}S_{\perp 
Y}^*\rangle)$ where $\langle AB^*\rangle\equiv \sum \exp 
(i\bf{Qr_j})$$A(0)B^*(\bf{r_j})$, $\bf{S}_\perp =\bf{Q}\times 
(\bf{S}\times \bf{Q})$, $\bf{S}$ is the spin at the position $\bf{r_j}$, 
and the coordinate axes $X$ and $Z$ run along $\bf{Q}$ and normal to the 
scattering plane, respectively. It depends on the helicity of the spiral 
and on the angle between $\bf{e}$ and $\bf{Q}$: the difference between the 
channels is maximal for $\bf{e}||\bf{Q}$, and is absent for 
$\bf{e}\perp\bf{Q}$. Fig. 4(a) shows scans through the (1/2,-1/2,1/2) 
$\pm$$\tau_1$ peaks in the ($h$,-$h$,$l$) plane in the single-domain state 
of Fig. 3(b). The direction of this scan is shown in Fig. 3(b) with dashed 
line. In this geometry, the intensity ratio of the +$\tau_1$ and -$\tau_1$ 
peaks, calculated for the spiral rotating in the plane of $\bf{P}$ and 
$\tau_1$, is $\sim$8. The data of Fig. 4(a) fit well to the sum of these 
peaks with the intensity ratio of 9, and a weak central peak, possibly 
coming from $\lambda$/2 contamination. Note that for any non-chiral 
structure, such as a spin density wave, or for a sample equally populated 
with spirals of the both helicities, both the $\tau_1$ peaks should be of 
the same intensity. Fig. 4(b) shows scans through a different magnetic 
peak, (1/2,1/2,1/2) $\pm$$\bf{\tau}$, in the ($h$,2$l$-$h$,$l$) scattering 
plane in the two-domain state of Fig. 3(a). As discussed in Ref. 
\cite{Lee}, the $\tau_1$ and $\tau_2$ domains produce peaks in the 
$\pm$$\delta$/2 and $\pm$$\delta$ positions, respectively, in these scans, 
giving rise to the broad peaks of Fig. 4(b). In this geometry, $\bf{e}$ is 
normal to $\bf{Q}$, and no difference between the (+-) and (-+) channels 
is observed, as expected. Thus, our data provide strong evidence for the 
spiral model discussed in the literature.\cite{Sos1} Moreover, we show 
that magnetic domains are fully chiral (contain a single-helicity spiral 
for a given $\bf{\tau}$), as expected on symmetry grounds.

Spontaneous formation of large single crystals containing a single 
ferroelastic, ferroelectric, and (upon application of an electric field) 
single-$\bf{\tau}$ chiral magnetic domain, while seemingly surprising, was 
achieved by design. B$_2$O$_3$ was added to the flux to reduce the 
liquidus temperature down to $\sim$620 ${}^\circ$C, thus allowing the 
crystals to grow below $T_C$ in the polar state.\cite{Kubel} Effectively, 
the system behaves as a pyroelectric, and formation of a unipolar 
ferroelectric domain is favored. While the growth temperature was not 
measured directly, we found that annealing the crystals at 
$T$$\approx$$T_C$ produces ferroelastic domains, strongly supporting this 
mechanism. The origin of the chiral magnetic state lies in the symmetry of 
the system, which fixes the helicity of the spiral once $\bf{P}$ and 
$\bf{\tau}$ are chosen, as observed in our experiments.

The above data demonstrate that high level of control of electric and 
magnetic properties can be achieved in single-crystal BFO. The samples 
contain a single ferroelectric domain, reversible ferroelectric switching 
occurs between only two well defined polarization states, and a 
single-$\tau$ magnetic chiral domain can be stabilized by piezoelectric 
strain. Note that 50-$\mu$m-thick single-domain crystals were reported in 
Refs. \cite{Lebeugle,Kubel} , but no reversible control of $\bf{P}$ and no 
control of the equivalent magnetic domains was achieved, and the chirality 
was not measured. Also, as shown previously, the magnetic structure 
rotates together with $\bf{P}$ when an electric field is 
applied.\cite{Lee,Lebeugle} Such level of control is currently 
unachievable in thin films which consist of a patchwork of ferroelastic 
microdomains, with domain-dependent properties in an applied 
field.\cite{Zhao,Chu} Single-crystal BFO is therefore suitable for studies 
of intrinsic properties of this multiferroic, as well as for fabrication 
of devices with precisely controlled properties. Our results may also help 
identify prospective approaches for growth of uniform thin films. Finally, 
BFO provides an intriguing example of successful ``chiral 
resolution''\cite{Jacques}, in which a large single crystal is prepared in 
a chiral magnetic state. This result reveals the symmetries involved in 
the coupling between the magnetic and electric order parameters in BFO. 
Chirality and the inversion symmetry are strongly coupled in BFO, as they 
are in many other multiferroics, such as TbMnO$_3$, in which the chirality 
is also coupled to an electric field.\cite{Tokura} Functional properties 
of these materials are defined by the underlying symmetry, and 
understanding the symmetry is essential for control and application of 
these properties.

In summary, we report preparation of mm-sized single crystals of 
BFO which grow below the Curie temperature and consist of a single 
ferroelectric domain. The electric polarization can be switched by an 
electric field applied along the cubic direction, and only 71${}^\circ$ 
rotation of $\bf{P}$ is observed. Magnetic state in BFO is 
chiral, and for a given wavevector the spiral is of a single helicity 
defined by the symmetry of the system via the direction of 
$\bf{P}$. The equivalent spiral domains can be reversibly 
controlled by an electric field, and a chiral magnetic state with a 
single-$\bf{\tau}$ spiral can be stabilized by field-induced strain in 
the entire crystal. This is a highly unusual way of achieving chiral 
resolution and of controlling the magnetic state. These properties 
of single-crystal BFO make it a promising system for multiferroics 
research and experimental device fabrication.

This work was supported by the NSF under Grants No. DMR-0704487, 
and DMR-0520471.

\begin{figure}[htb]
\begin{center}
\caption{(Color online) (a) Possible electric polarization directions in 
the pseudocubic cell of BiFeO$_3$. (b) Equivalent magnetic spirals in a 
single ferroelectric domain. (c) A single ferroelectric/magnetic domain 
state. (d) Switching of the ferroelectric domains in an applied field. (e) 
and (f) show the orientation of $E$ with respect to $\tau_1$, $\tau_2$, 
and $\tau_3$, and its projection $E_{in}$ on the plane containing these 
vectors. For simplicity, only one antiferromagnetic sublattice in the 
spirals is shown, and the modulation wavelength is decreased.}
\end{center}
\end{figure}

\begin{figure}[htb] 
\begin{center} 
\caption{(Color online) Representative AFM (a) and PFM (b,c) images of the 
(001) surface. Out-of-plane (d), and in-plane (e,f) PFM images of a 
ferroelectric domain created by a biased PFM tip. Arrows indicate the 
polarization direction probed in each image. The polarization vectors in 
the ferroelectric domains are shown schematically in (g).}
\end{center} 
\end{figure}

\begin{figure}[htb]
\begin{center}
\caption{(Color online) Magnetic peaks near the (1/2,-1/2,1/2) position in 
the ($h$,-$h$,$l$) plane for $E$=0 in the as-prepared sample (a), for 
$E$=1.3 MV/m (b), and for $E$=-1.3 MV/m (c). The ellipse in (b) shows the 
calculated instrumental resolution. Crosses indicate projections of the 
$\tau_1$, $\tau_2$, and $\tau_3$ peaks on the scattering plane. Circles 
show the positions of the peaks used in the fits described in the text. 
The fits and the residuals are shown to the right of the main panels. 
Populations of the magnetic domains $\tau_1$ and $\tau_2$ determined in 
these fits are shown in (d) as functions of $E$.}
\end{center}
\end{figure}

\begin{figure}[htb]
\begin{center}
\caption{(Color online) Polarized neutron scans in the (+-) and (-+) spin 
flip channels. (a) ($h$,-$h$,1.5-2$h$) scan through domain $\tau_1$ near 
the (1/2,-1/2,1/2) position in the ($h$,-$h$,$l$) plane. The direction of 
this scan is shown in Fig. 3(b) with dashed line. Error bars are from
counting statistics. (b) 
(0.5+$\Delta$$h$,0.5-$\Delta$$h$,0.5)  scan through domains $\tau_1$ and 
$\tau_2$ near the (1/2,1/2,1/2) position in the ($h$,2$l$-$h$,$l$) plane. 
Arrows show the positions of the peaks due to these domains. Insets show 
the orientation of the scattering vector $\bf{Q}$ with respect to the 
plane of the magnetic spiral, as well as vectors $\bf{e}$, $\bf{P}$, and 
$\bf{\tau}$.}
\end{center}
\end{figure}

\end{document}